# Coupling constant $g_{a_0\omega\gamma}$ as derived from light cone QCD sum rules

C Aydin[*] and A H Yilmaz[†]

*Physics Department, Karadeniz Technical University, 61080 Trabzon, Turkey*
[*] coskun@ktu.edu.tr
[†] hakany@ktu.edu.tr

**Abstract**

We investigate the $a_0\omega\gamma$ vertex and estimate the coupling constant $g_{a_0\omega\gamma}$ in the framework of the light cone QCD sum rules. We compare our result with the value of the coupling constant was calculated in the QCD sum rules.

A QCD sum rule is an analytical method designed to provide an approximate calculation scheme for strong coupling QCD and in particular to account for non-perturbative effects. QCD sum rule has been a widely used working tool in hadron phenomenology [1]. The method has many advantages. Instead of a model-dependent treatment in terms of constituent quarks, hadrons are represented by their interpolating quark currents taken at large virtualities. The correlation function of these currents is introduced and treated in the framework of the operator product expansion (OPE), where the short and long-distance quark-gluon interactions are separated. The former are calculated using QCD perturbation theory, whereas the latter are parametrized in terms of universal vacuum condansates or light-cone distrubution amplitudes. The QCD light-cone sum rule method (LCSR) has been suggested in [2-4] and is a combination of the operator product expansion (OPE) on the light-cone [5-7] with QCD sum rule techniques [8]. For a review of the method and results we refer to Refs.[9,10].

Radiative transitions between pseudoscalar (P) and vector (V) mesons have been an important subject in low-energy hadron physics for more than three decades. These transitions have been considered as phenomenological quark models, potential models, bag models, and effective Lagrangian methods [11,12]. Since they determine the strength of the hadron interactions, among the characteristics of the strong interaction processes $g_{PV\gamma}$ coupling constant plays one of the most important roles,. In the quark models, V → P+γ decays (V = φ, ρ, ω) ; P = π, η, η′) are reduced by the quark magnetic moment with transition s = 1 → s = 0, where s is the total spin of the $q\bar{q}$ − system (in the corresponding meson). These quantities can be calculated directly from QCD. Low-energy hadron interactions are governed by nonperturbative QCD so that it is very difficult to get the numerical values of the coupling constants from the first principles. For that reason a semiphenomenological method of QCD sum rules can be used, which nowadays is the standart tool for studying of various characteristics of hadron interactions. On the other hand, vector meson-pseudoscalar meson-photon VPγ–vertex also plays a role in photoproduction reactions of vector mesons on nucleons. It should be notable that in these decays (V→Pγ) the four-momentum of the pseudoscalar meson P is time-like, $P'^2 > 0$, whereas in the pseudoscalar exchange amplitude contributing to the photoproduction of vector mesons it is space-like $P'^2 < 0$. Therefore, it is of interest to study the effective coupling constant $g_{VP\gamma}$ from another point of view as well. In addition, the same model predicts amplitudes for energetically allowed $S \to V\gamma$ processes, for example $f_0 \to \omega\gamma$, $f_0 \to \rho\gamma$, $a_0 \to \omega\gamma$, and $a_0 \to \rho^0\gamma$ etc. The coupling constant of $a_0 \to \omega\gamma$ and $a_0 \to \rho\gamma$ decays were calculated [13] in the tree-point QCD sum rules and obtained as $g_{a_0\omega\gamma} = 0.75$ and $g_{a_0\rho\gamma} = 2.00$, respectively. From the experimental point of view, $a_0(980)$ is well established, but the nature and the quark substructure of this scalar meson, the question whether they are conventional $q\bar{q}$ states states has been a subject of controversy. On the other hand, they are relevant hadronic degrees of freedom, and therefore the role they play in hadronic processes should also be studied besides the questions of theire nature. In this work, we calculated the coupling constant $g_{a_0\omega\gamma}$ by applying light cone QCD sum rules, which provide an efficient and model-independent method to study many hadronic observables, such as decay constants and form factors in terms of non-perturbative contributions proportional to the quark and gluon condensates [1, 14-16].

In order to derive the light cone QCD sum rule for the coupling constants $g_{a_0\omega\gamma}$, we consider the following two point correlation function

$$T_\nu(p, p') = i\int d^4x e^{ip'\cdot x} \langle 0|T\{j_\nu^\omega(x) j_{a_0}(0)\}|0\rangle_\gamma, \qquad (1)$$

where $\gamma$ denotes the external electromagnetic field, and $j_\nu^\omega$ and $j_{a_0}$ and are the interpolating current for the $\omega$ meson and $a_0$, respectively. $j_\nu^\omega = \frac{1}{2}\left(\bar{u}^a\gamma_\nu u^a + \bar{d}^a\gamma_\nu d^a\right)$ and $j_{a_0} = \frac{1}{2}\left(\bar{u}^b u^b - \bar{d}^a d^b\right)$ are the interpolating currents for $\omega$ and $a_0$ meson with $u$ and $d$ denoting up and down quark field, respectively. Where $a$ and $b$ are the color indices.

We therefore sature the dispersion relation satisfied by the vertex function $T_\mu$ by these lowest lying meson states in the vector and the scalar channels, and in this way we obtain for the physical part at the phenomenological level the Eq. (1) can be expressed as

$$T_\nu(p, p') = \frac{\langle 0|j_\nu^\omega|\omega\rangle\langle\omega(p)|a_0(p')\rangle_\gamma\langle a_0|j_{a_0}|0\rangle}{(p^2 - m_\omega^2)(p'^2 - m_{a_0}^2)}. \qquad (2)$$

The overlap amplitude $\lambda_{a_0} = \langle a_0|j_{a_0}|0\rangle$ of the $a_0$-meson in this expression has been determined in Ref. [13]. On the other hand the overlap amplitude $\lambda_\omega$ of the vector meson is explained by $\langle 0|j_\nu^\omega|\omega\rangle = \lambda_\omega u_\nu$, where $u_\nu$ is polarization vector of $\omega$. The matrix element of the electromagnetic current is given by the polarization vector of $\omega$,

$$\langle \omega(p) | a_0(p') \rangle_\gamma = -i \frac{e}{m_\omega} g_{a_0 \omega \gamma} K(q^2)(p.q u_\nu - u.q p_\nu) \varepsilon_\nu, \tag{3}$$

where $q = p - p'$, $\varepsilon$ is the polarization of the photon, and $K(q^2)$ is a form factor with $K(0) = 1$. This expression defines the coupling constant through the effective Lagrangian

$$\mathcal{L} = \frac{e}{m_\omega} g_{a_0 \omega \gamma} \partial^\alpha \omega^\beta (\partial_\alpha A_\beta - \partial_\beta A_\alpha) a_0, \tag{4}$$

describing the $a_0 \omega \gamma$-vertex.

In this calculation the full light quark propagator with both perturbative and nonperturbative contribution is used, and it is given as [17]

$$iS(x,0) = \langle 0 | T\{\bar{q}(x) q(0)\} | 0 \rangle$$

$$= i \frac{\not{x}}{2\pi^2 x^4} - \frac{\langle \bar{q} q \rangle}{12} - \frac{x^2}{192} m_0^2 \langle \bar{q} q \rangle - i g_s \frac{1}{16\pi^2} \int_0^1 du \left\{ \frac{\not{x}}{x^2} \sigma_{\mu\nu} G^{\mu\nu}(ux) - 4iu \frac{x_\mu}{x^2} G^{\mu\nu}(ux) \gamma_\nu \right\} + ... \tag{5}$$

where the terms proportional to light quark mass $m_u$ or $m_d$ are neglected. After a straightforward computation we have

$$T_\mu(p,q) = 2i \int d^4 x e^{ipx} A(x_\rho g_{\mu\tau} - x_\tau g_{\mu\rho}) \langle \gamma(q) | \bar{q}(x) \sigma_{\tau\rho} q(0) | 0 \rangle \tag{6}$$

where $A = \frac{i}{2\pi^2 x^4}$, and higher twist corrections are neglected since they are known to make a small contribution [18]. In order to evaluate the two point correlation function further, we need the matrix elements $\langle \gamma(q) | \bar{q}(x) \sigma_{\tau\rho} q(0) | 0 \rangle$. This matrix element can be expanded in the light cone photon wave function [19,20]

$$\langle \gamma(q) | \bar{q} \sigma_{\alpha\beta} q | 0 \rangle = i e_q \langle \bar{q} q \rangle \int_0^1 du e^{iuqx}$$

$$\times \{ (\varepsilon_\alpha q_\beta - \varepsilon_\beta q_\alpha)[\chi \varphi(u) + x^2[g_1(u) - g_2(u)]] + [q.x(\varepsilon_\alpha x_\beta - \varepsilon_\beta x_\alpha) + \varepsilon.x(x_\alpha q_\beta - x_\beta q_\alpha)] g_2(u) \} \tag{7}$$

In these equations $e_q$ is the corresponding quark charge. $\chi$ is the magnetic susceptibility, $\varphi(u)$ is leading twist two and $g_1(u)$ and $g_2(u)$ are the twist four photon wave functions. The main difference between the tradiatonal QCD sum rules and light cone QCD sum rule is the appearance of these wave function. Light cone QCD sum rules corresponds to summation of an infinite set of terms in the expansion of this matrix element on the tradiational sum rules. The price one pays for this is the appearence of a priori unknown photon wave functions.

After evaluating the Fourier transform for the M1 structure and then performing the double Borel transformation with respect to the variables $Q_1^2 = -p_1^2$ and $Q_2^2 = -p_2^2$, we finally obtain the following sum rule for the coupling constant $g_{a_0\omega\gamma}$

$$g_{a_0\omega\gamma} = \frac{(e_u - e_d)m_\omega <\bar{u}u>}{2\lambda_{a_0}\lambda_\omega} e^{m_\omega^2/M_1^2} e^{m_{a_0}^2/M_2^2} \left\{ -M^2 \chi \phi(u_0) E_0(s_0/M^2) + 4g_1(u_0) \right\} \quad (8)$$

where the function

$$E_0(s_0/M^2) = 1 - e^{-s_0/M^2} \quad (9)$$

is the factor used to subtract the continuum, $s_0$ being the continuum threshold, and

$$u_0 = \frac{M_2^2}{M_1^2 + M_2^2}, \quad M^2 = \frac{M_1^2 M_2^2}{M_1^2 + M_2^2} \quad (10)$$

with $M_1^2$ and $M_2^2$ are the Borel parameters in the $\omega$ and $a_0$ channels.

For the numerical evaluation of the sum rule we used the values $<\bar{u}u> = -0.014\,\text{GeV}^3$, $m_{a_0} = 0.98\,\text{GeV}$, $\lambda_{a_0} = 0.21 \pm 0.05\,\text{GeV}^2$, $m_\omega = 0.782\,\text{GeV}$ [13], and for the magnetic susceptibility $\chi = -3.15\,\text{GeV}^{-2}$ [16]. We note that neglecting the electron mass the $e^+e^-$ decay width of $\omega$ meson is given as $\Gamma(\omega \to e^+e^-) = \frac{4\pi\alpha^2}{3}\left(\frac{\lambda_\omega}{3}\right)^2$. Then using the value from the experimental leptonic decay width $\Gamma(\omega \to e^+e^-) = 0.60 \pm 0.02$ of keV for $\omega$ [13], we obtain the value $\lambda_\omega = (0.108 \pm 0.002)\,\text{GeV}^2$ for the overlap amplitude $\omega$ meson. Using the conformal invariance of QCD up to one loop order, the photon wave functions can be expanded in terms of Gegenbauer polynomials; each term corresponding to contributions from operators of various conformal spin. Due to conformal invariance of QCD up to one loop, each term in this expansion is renormalized separately and the form of these wave functions at a sufficiently high scale is well known. In [19] it is shown that even at small scales, the wave functions do not deviate considerably from their asymptotic form and hence we will use the asymptotic forms of the photon wave function given by:

$$\varphi(u_0) = 6u_0(1-u_0),$$

$$g_1(u_0) = -\frac{1}{8}(1-u_0)(1-3u_0), \quad (11)$$

$$g_2(u_0) = -\frac{1}{4}(1-u_0)^2$$

Since $m_{a_0} \approx m_\omega$, we will set $M_1^2 = M_2^2 = 2M^2$ which sets $u_0 = 1/2$. Note that in this approximation, we only need the value of the wave functions at a single point; namely at $u_0 = 1/2$ and hence the functional forms of the photon wave functions are not relevant.

In Fig. 1 we show the dependence of the coupling constant $g_{a_0\omega\gamma}$ on parameter $M^2$ at some different values of the continuum threshold as $s_0 = 1.8$, 2.0 and 2.2 GeV$^2$. Since the Borel masses $M_1^2$ and $M_2^2$ are the auxiliary parameters and the physical quantities should not depend on them, one must look for the region where $g_{a_0\omega\gamma}$ is practically independent of $M_1^2$ and $M_2^2$. We determined that this condition is satisfied in the interval $1.0\ GeV^2 \leq M^2 \leq 1.4\ GeV^2$. The variation of the coupling constant $g_{a_0\omega\gamma}$ as a function of different values $M^2$ and $s_0$ are shown in Fig. 2. Examination of this figure points out that the sum rule is rather stable with these reasonable variations of $M^2$.

Within the range of variation of the parameters we used, we found that the leading twist wave function gives the main contribution. Its contribution constitutes more than 90% of the final result. Hence we conclude that the expansion of the coupling constant in terms of contributions of increasing twist converges very rapidly, and thus our result is reliable. We choose the middle value $M^2 = 1.2$ GeV$^2$ for the Borel parameter in its interval of variation and obtain the coupling constant $g_{a_0\omega\gamma}$ as $g_{a_0\omega\gamma} = 2.57 \pm 0.21$. This result is agreed with the work of Black et all.[20]. If we start with $j_\nu^\omega = \frac{1}{6}\left(\bar{u}^a \gamma_\nu u^a + \bar{d}^a \gamma_\nu d^a\right)$, we have $g_{a_0\omega\gamma}$ as $g_{a_0\omega\gamma} = 0.85 \pm 0.21$ which is very close to the result obtained in Gokalp and Yilmaz's work [13].

The error quoted above include the variations of the results with respect to the Borel parameter, $M^2$, the continuum threshold, $s_0$, and the uncertainties in the photon wave functions used.

In this work we have only nonperturbative contribution since the structure of the correlator is such that perturbation contribution to it is absent due to the odd number of gamma matrix and massles quark included.

**Acknowledgments**

We are grateful to Dr.Altug Ozpineci and Prof. T.M. Aliev for careful reading of the manuscript and helpful suggestions about its revision.

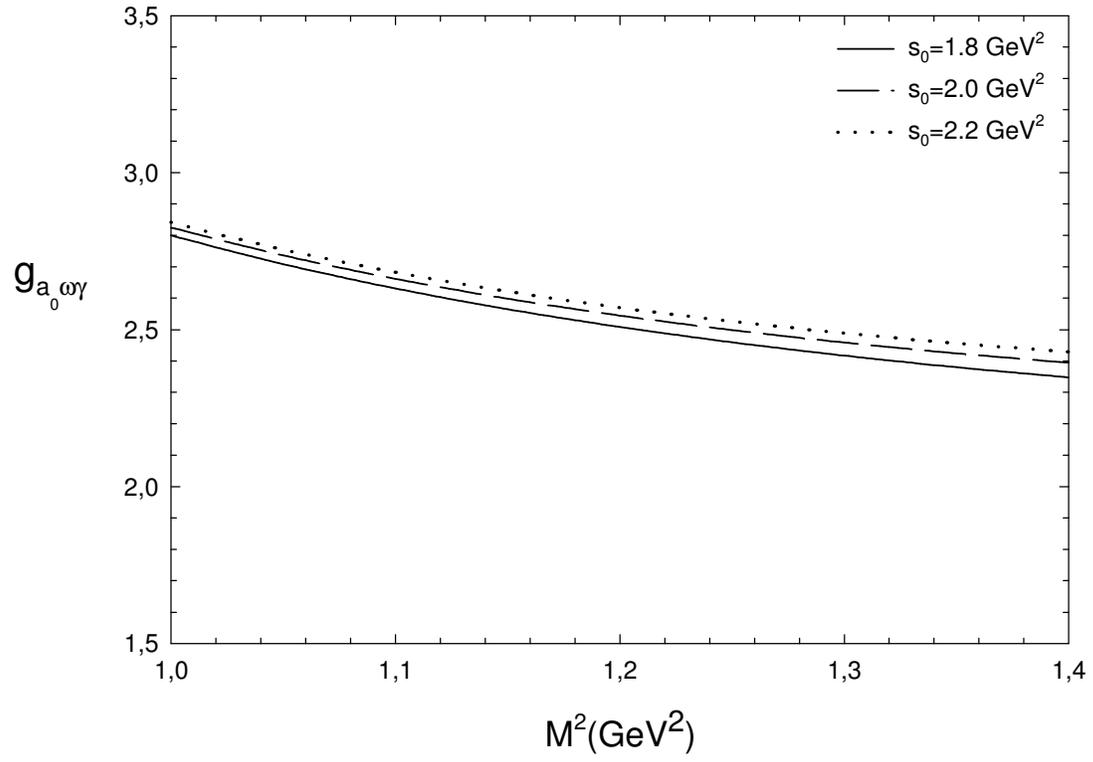

Figure.1. The coupling constant $g_{a_0\omega\gamma}$ as a function of $M^2$ for different values of the threshold parameters $s_0$.

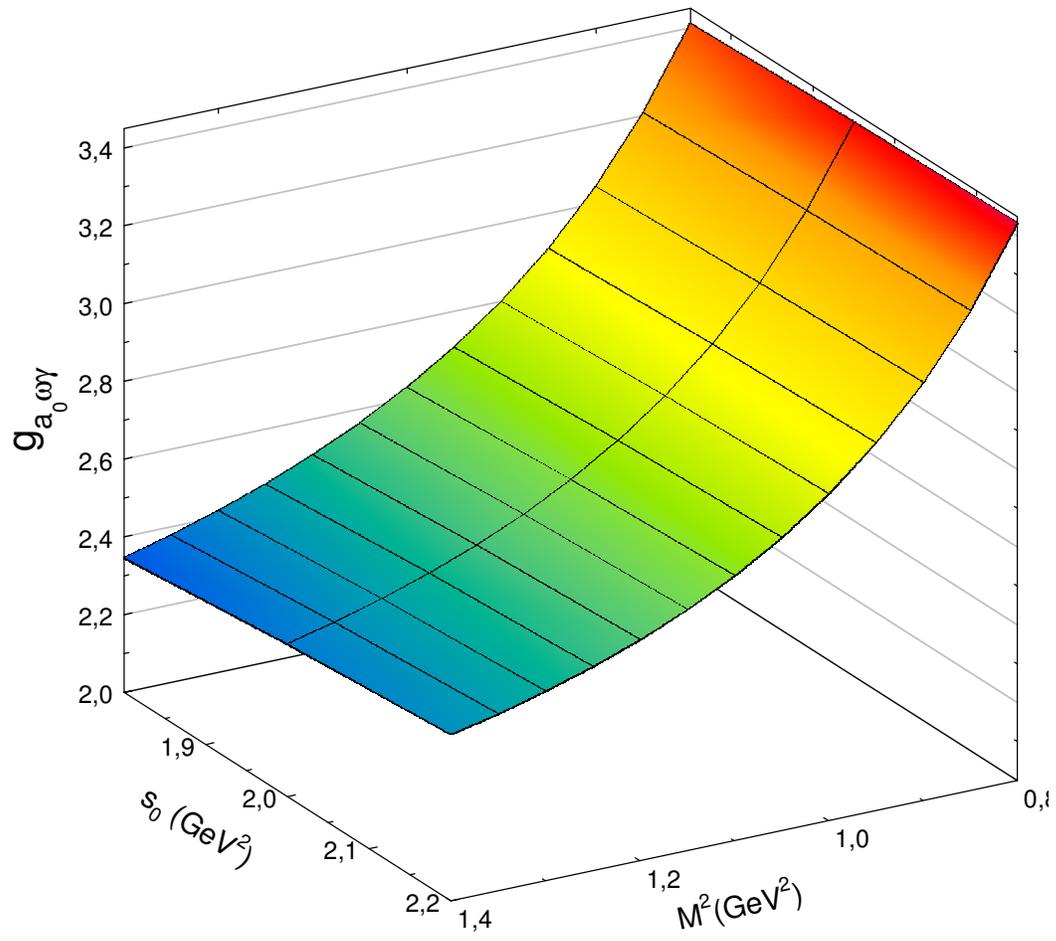

Figure 2. Coupling constant $g_{a_0\omega\gamma}$ as a function of $M^2$ and $s_0$.